# Phase transition and anomalous electronic behavior in layered dichalcogenide CuS (covellite) probed by NQR

R.R. Gainov [1*], A.V. Dooglav [1], I.N. Pen'kov [2], I.R. Mukhamedshin [1,3], N.N. Mozgova [4], A.V. Evlampiev [1], and I.A. Bryzgalov [5]

[1] *Department of Physics, Magnetic RadioSpectroscopy Laboratory, Kazan State University, Kazan, Kremlevskaya str. 18, 420008, Russian Federation*

[2] *Department of Geology, Kazan State University, Kazan, Kremlevskaya str. 4/5, 420111, Russian Federation*

[3] *Laboratoire de Physique des Solides, UMR 8502, Universite Paris-Sud, 91405 Orsay, France*

[4] *Institute of geology of ore deposits, petrography, mineralogy and geochemistry (Russian Academy of Science), Staromonetny per. 35, Moscow 109017, Russian Federation*

[5] *Department of Geology, Moscow State University, Moscow, Vorob'evy gory, 119991, Russian Federation*

\* Corresponding author: tel.: 7-843-2315175; fax.: 7-843-2387201.
Electronic address: ramil.gainov@ksu.ru (R.R. Gainov).

Nuclear quadrupole resonance (NQR) on copper nuclei has been applied for studies of the electronic properties of quasi-two-dimensional low-temperature superconductor CuS (covellite) in the temperature region between 1.47 and 290 K. Two NQR signals corresponding to two non-equivalent sites of copper in the structure, Cu(1) and Cu(2), has been found. The temperature dependences of copper quadrupole frequencies, line-widths and spin-lattice relaxation rates, which so far had never been investigated so precisely for this material, altogether demonstrate the structural phase transition near 55 K, which accompanies transformations of electronic spectrum not typical for simple metals. The analysis of NQR results and their comparison with literature data show that the valence of copper ions at both sites is intermediate in character between monovalent and divalent states with the dominant of the former. It has been found that there is a strong hybridization of Cu(1) and Cu(2) conduction bands at low temperatures, indicating that the charge delocalization between these ions takes place even in 2D regime. Based on our data, the occurrence of energy gap, charge fluctuations and charge-density waves, as well as the nature of phase transition in CuS are discussed. It is concluded that some physical properties of CuS are similar to those of high-temperature superconductors (HTSC) in normal state.

## I. INTRODUCTION

Metallic sulfides exhibit a fascinating variety in crystal chemistry as well as in physical properties, with both scientific interest and practical applications.[1-4] The role of binary copper chalcogenide with chemical composition CuS (referred to as covellite or covelline) should be specially emphasized. The pure and doped synthetic analogs of CuS have the practical application in optical, photovoltaic and solar devices[5] (i); this compound has been extensively studied owing to its possible applications in the synthesis of composite high-temperature superconductors (HTSC)[6] (ii); and as cathode material in lithium rechargeable batteries[7] (iii); CuS has been included also in nanotechnological investigations[8] (iv). Furthermore, (v) since covellite represents one of the wide-spread specimen of industrial copper-ore deposits, for more efficient extraction and working up of copper the different phase-diagnostic characteristics and magnetic properties of this material among others are essential.[3] From scientific point of view, covellite has drawn significant interest as a superconductor. It is known that both synthetic[9,10] and natural[11] CuS shows excellent metal conductivity and condense into the superconducting state below about 1.6 K. One of the pronounced features of CuS is the occurrence of quasi-two-dimensional, layer-type structure, which is built up of the triangular units $CuS_3$ (Fig. 1), similar to some extent to those in Cu-oxide HTSC's (constructed from $CuO_4$-layers). Since after discovery of HTSC[12] the nature of superconductivity (SC) in these materials is still unclear up to now, some researchers suggest the studying of relationship of mechanisms of SC through an analogy between Cu-O and either Co-O bonds in Co-oxide HTSC's[13] or Cu-S bonds in CuS.[14,15] In addition, it has been found that HTSC materials in the normal state and transition metal dichalcogenides ($CuS_2$, $NbS_2$, $VSe_2$, $TaS_2$, $Mo_xW_{1-x}Se_2$, and other) exhibit some general and difficult for explanation electronic features (energy gaps, charge-density waves), which will be helpful to study on the base of common viewpoints.[16-20] Finally, 2D transition metal CuS by itself possesses a set of indistinguishable peculiarities of crystal-chemistry, in particular: indecisive valence state of copper,[21-23] the character of low-temperature phase transition,[24,25] anomalous negative Knight shift,[15] and other. In any case, the clarification of problems mentioned requires the study of local properties of covellite, for which the nuclear-resonance methods are most suitable.

In this paper, we report the results of nuclear quadrupole resonance (NQR) measurements including the $T$-dependences of $^{63}Cu$ quadrupole frequencies, line-widths, as well as $^{63}Cu$ spin-lattice relaxation $T_1$ in series of covellite samples with formal compositions $Cu_{1+x}S$ (0≤$x$≤0.30) in the temperature range 1.47–290 K. The single $^{63}Cu$ NQR signal in stoichiometric CuS was first observed by Abdullin[26] at about 14.88 MHz, but only at 4.2 and 77 K. Later Itoh and Tnabe extended this study by measurements of $T$-dependences of NQR frequency $v_Q$[27,28] and nuclear relaxation $T_1$[27] above 4.2 K. In addition, from NMR spectra Itoh[27] supposed the existence of another $^{63}Cu$ NQR signal at about 1.5 MHz above the temperature of structural phase transition at $T_{PT}$ = 55 K. In contrast, Saito,[15] also from NMR spectra, predicted the occurrence of this NQR signal not only above, but also below $T_{PT}$ at about 1.8 MHz. In order to specify the low-temperature features of covellite electronic structure and resolve the opposing data, we reexamined thoroughly the $T$-dependences of copper spectroscopic and relaxation parameters, especially at low $T$. In particular, below $T_{PT}$ we have found experimentally the low-frequency $^{63}Cu$ NQR signal at 1.87 MHz (4.2 K) and have studied its nuclear relaxation $T_1(T)$; we have revealed the change of the character of $^{63}Cu$ NQR line-width $dv_Q$ broadening of high-frequency signal and rapid divergence of $^{63}T_1 \cdot T$ from constant behavior, typical for most metals. On the basis of obtained experimental data, some aspects of covellite crystal structure and its physical properties are presented and discussed.

The paper is organized as follows. Since data concerning CuS are numerous, but rather odd, often inconsistent and not systematized, in Section II we briefly describe crystal structure and related important electronic properties. Experimental procedures, including the characterization of the samples, are given in Section III. The next Section (IV) contains necessary NMR-NQR background. In Section V we report our results with subsequent discussions in Section VI. The paper concludes with a summary in Section VII.

## II. CRYSTAL-CHEMISTRY AND ELECTRONIC PROPERTIES

The crystal structure of covellite CuS is described in numerous papers.[21,29-35] The elementary cell of covellite at room-$T$ has a hexagonal symmetry corresponding to a space group $P6_3/mmc$ with 6 formula units per unit cell. The crystal structure of CuS can be presented as "sandwich" or packet, which consists of three alternating layers $A_1$-B-$A_2$ (Fig. 1). The layers $A_1$ and $A_2$ are made up of $CuS_4$-tetrahedra, jointed by vertexes. The layer B represents the net of $CuS_3$-triangles, combined by apexes. In frame of packet the $CuS_4$-tetrahedra of layers $A_1$ and $A_2$ are turned in opposite directions. Usually, three sulfur atoms of $CuS_3$-triangle are referred to as S(1), one of them is common with $CuS_4$-tetrahedra (Fig. 1). Other sulfur atoms of $CuS_4$-tetrahedra are referred to as S(2). Copper atoms in triangular and tetrahedral coordination are marked as Cu(1) and Cu(2), respectively. The packets $A_1$-B-$A_2$ are connected together along the $c$-axis by S(2)-S(2) bonds ("dumbbells").

Covellite CuS at room-$T$ represents the metal with hole conductivity,[4] which becomes superconducting below $T_C$ ranging from 1.72 K[15] down to 1.31 K.[10] Recent studies show that CuS is attributed to Class I SC materials[11] and, more likely, CuS might be an anisotropic SC.[36] It is interesting that covellite CuS is the first specimen of natural solids (i.e. minerals), in which SC transition was detected.[11]

In Table I, the interatomic distances and angles between chemical bonds in CuS are reported. Crystal-chemists noted that trigonal Cu(1)-S(1) bonds (about 2.19 Å) are strikingly shorter than Cu-S bonds in triangular units in the most of other copper sulfides (about 2.33 Å).[21,33] This fact strongly suggests that the Cu(l)-S(l) bond strength should be very large, and the layer formed by Cu(l) and S(l) should be very strong.[33] On the contrary, the bond length in the Cu(2) tetrahedron is 2.305-2.335, slightly larger than the average value of tetrahedral Cu-S bond lengths: 2.302(l) Å in chalcopyrite $CuFeS_2$ and 2.305 Å (average) in cubanite $CuFe_2S_3$.[21] In addition, it was found that triangular coordinated Cu(l) ions have a large thermal motion along the $c$-axis (i.e. perpendicular to the plane of Cu(1)S$_3$-units), whereas the tetrahedrally coordinated Cu(2) ion motion is practically isotropic.[21]

However, situation changes with decreasing temperature. The heat capacity studies showed that the anomaly is observed at temperatures around 55 K.[37] It was found that heat capacity is proportional to $T^2$ in the region 5-20 K. Researchers explained the $T^2$-law by Debye theory of low-frequency modes of lattice vibrations with the assumption of a quasi-two-dimensional (2D) lattice, which, probably, is caused by weak bonding in the $c$-axis direction. At higher temperatures ($T$>50 K) the lattice heat capacity was estimated in terms of a three-dimensional (3D) model. The gradual transformation from 3D to 2D lattice of CuS with decreasing $T$ is indirectly confirmed by theoretical calculations[25] and by experimental studies of electrical conductivity,[15] and $^{63}$Cu spin-lattice relaxation[27] in oriented powder samples.

The temperature dependence of X-ray and neutron diffraction revealed the gradual anomalous contraction of the unit cell of CuS below 90 K with subsequent second-order phase transition (PT) at about $T_{PT}$ = 55 K to a phase with, more likely, orthorhombic symmetry, space group *Cmcm*.[24] Later the occurrence of second-order PT near $T_{PT}$ was reexamined by other X-ray diffraction studies[38] and also observed through the electric resistivity,[14] Hall coefficient,[14] and Raman spectroscopy[39] measurement. This PT can be visualized as a shift of the layers formed by CuS$_3$-triangles with respect to the CuS$_4$-layers perpendicular to the $c$-axis with the change of the Cu(2)-S(1)-Cu(2) bonding angle from 180º at room temperature to 170º below $T_{PT}$ (Fig. 1).[24] Furthermore, this distortion involves the slight, but important changes of Cu(1) and Cu(2) bond lengths (Table I). It was assumed originally that formation of metal Cu(1)-Cu(2) bond (3.04 Å) is a driving force for PT,[24] however theoreticians proposed that PT is caused by van der Waals interactions of the S(1)-S(2) contacts (3.64 Å) of adjacent CuS$_3$ and CuS$_4$ units.[25] Another possibility is that a change in the bonding nature of the S(2)-S(2) or Cu(2)-S(1) pairs on varying $T$ may stimulate the PT.[38]

Another important problem still lacking of clarity is the active valence states of copper and sulfur, their distribution in the crystal structure of CuS. In particular, the $(Cu^{2+})(S_2^{2-})(Cu_2^{1+})(S^{2-})$ valence formalism was proposed,[30,31,40-42] i.e. there exist both valence forms of copper. However, EPR studies showed no signal of paramagnetic $Cu^{2+}$ at room and low temperatures in CuS.[23] Some measurements by X-ray photoelectron spectroscopy (XPS) and X-ray absorption spectroscopy (XAS)[43-49] point out that covellite CuS is better described as compound consisting of only $Cu^{1+}$. Other XPS,[44,50] X-ray emission spectroscopy (XES)[50] and X-ray absorption near-edge spectroscopy XANES[51] studies revealed the presence of two different electronic states of the S atoms; the S(2) sulfur atoms in CuS form S$_2$-dimers. Therefore the bonding was described as $(Cu^{1+})_3(S_2^{-2})(S^{-1})$[24] or $(Cu^{1+})_3(S_2^{-1})(S^{-2})$[25] in terms of ionic model. At the same time, other crystal-chemical and theoretical studies[21-23] pointed that, probably, Cu possesses a valence state, which is intermediate between $Cu^{1+}$ and $Cu^{2+}$. Some magnetic susceptibility measurements of CuS show the existence of magnetic moment $\mu_{eff}$ of about 0.28·$\mu_\beta$.[24] However, the results of recent studies do not provide evidence for such behavior.[14,27]

The Hall coefficient studies[14] experimentally proved the previous suggestions,[45] according to which the excellent metallic conduction of CuS among 3d transition metal sulfides is due to electronic holes in valence band, mainly constituted from the 3p-orbitals of sulphur. Analogical data were obtained by other researchers.[25,48] More detailed information was deduced by fluorescent XES investigations,[52] according to which the band of CuS is composed of three parts, constituted mainly by Cu-3d, S-3p and, to a less extent, S-3s orbitals.

### III. MATERIALS AND EXPERIMENTAL METHODICS

Overall, we have studied a few samples, hereafter denoted with order numbers No.1-No.5. Samples No.1-No.4 are synthetic; these sulfides were prepared by solid-phase reaction method of high-purity elements Cu and S. Amounts of Cu and S were taken in relations, which correspond to compositions: stoichiometric covellite - $Cu_{1.00}S$ (sample No.1), non-stoichiometric Cu-rich covellites $Cu_{1.10}S$ (No.2), $Cu_{1.05}S$ (No.3) and $Cu_{1.30}S$ (No.4). The synthesis was carried out in a sealed quartz tubes, evacuated down to residual pressure $10^{-1}$ Pa. The regime of preparation was the following: the heating at 480ºC for one week, gradual cooling during 2 hours. All synthetic samples had the blue color; however the samples became darker with decreasing of Cu amount ($x$) as compared to S content. The sample No.5 is a natural covellite, originating from Bor copper-ore deposit (Serbia). This sample looked as large (up to 1.5 cm) crystallites in the form of thin plates and had playing indigo-blue color with varnish glitter.

The electron-probe microanalyses (EPMA), carrying out by scanning microscope Camebax SX-50 (accelerating voltage 15 kV, beam current 30 nA), confirmed the chemical compositions of synthetic samples and showed that the composition of sample No.5 is close to stoichiometric covellite $Cu_{0.99}S_{1.00}$.

The X-ray diffraction study confirmed the existence of covellite CuS phase in all samples. Moreover, it was found that non-stoichiometric samples (No.2-No.4) contain additional X-ray diffraction patterns, which correspond to another structural phase. The EPMA studies also pointed that non-stoichiometric samples (No.2-No.4) appear to be inhomogeneous and indicated the predominance of CuS phase. This phase heterogeneity is discussed in Section V and VI.

The copper nuclear resonance measurements were carried out using the standard home-built coherent pulsed NMR/NQR spectrometers. For better penetration of the high-frequency magnetic field all samples were crushed in an agate mortar to a particle size smaller than 75 μm and packed in epoxy resin "Stycast 1266". NQR spectra were taken "point by point" after the $\pi/2$-$\pi$ pulse sequence in the frequency sweep mode; after that, according to Fourier mapping algorithm,[53] detailed NQR spectra were created. The Cu nuclear spin-lattice (longitudinal) relaxation time $T_1$ was measured at the peak of the $^{63}$Cu NQR signal and calculated by plotting the Cu nuclear spin-echo intensity as a function of the time delay $\Delta t$ between a saturating and $\pi/2$-$\pi$ probing pulses.

### IV. NMR-NQR BACKGROUND

The NMR-NQR background will be reviewed only for the sake of clarity. The nuclear spins $I$ interact with their electronic environment through quadrupole (i.e. electric) and magnetic hyperfine couplings. In general, the nuclear magnetic resonance (NMR) spectrum of a quadrupole nucleus (i.e. nucleus with spin $I>½$) is described by the following spin Hamiltonian:

$$H = H_Z + H_Q + H_M. \tag{1}$$

Hamiltonian $H_Z$ is the Zeeman interaction of nuclear magnetic moments with the gyromagnetic ratio $\gamma_n$ with the applied external magnetic field $H_0$:

$$H_Z = -\gamma_n \hbar H_0 \cdot I \tag{2}$$

Hamiltonian $H_Q$ refers to as the coupling of nuclear quadrupole moment $eQ$ to the local crystal electric field gradient (EFG):

$$H_Q = \frac{eQV_{ZZ}}{4I(2I-1)}\{3I_Z^2 - I(I+1) + \frac{1}{2}\eta(I_+^2 + I_-^2)\} \tag{3}$$

with $V_{ZZ}$ the largest component of the crystal EFG tensor, $\eta=|V_{XX}-V_{YY}|/V_{ZZ}$ the asymmetry parameter showing the deviation of the EFG symmetry from the axial one, i.e. the value of $\eta$ lies in the range [0,1]. The EFG components satisfy Laplace equation: $V_{XX} + V_{YY} + V_{ZZ} = 0$.

Hamiltonian $H_M$ can be viewed as an interaction between the nuclear spin $I$ and a static plus time dependent local hyperfine field $F_L$ generated by the motion of conduction carriers. The static part of $F_L$ gives rise to a NMR shift expressed by the magnetic (Knight) shift $K = \Delta H/H_0$ due to the polarization of conduction charge spins. The fluctuating part of $F_L$ is the source of the nuclear spin-lattice relaxation.

The pure NQR spectrum is observed in the case of absence of the external ($H_0=0$) and internal ($H_{int} = 0$) static magnetic fields. The remaining $H_Q$ gives rise to doubly degenerate energy levels, between which NQR transitions are induced. The number of NQR lines is defined (i) by the amount of crystallographically nonequivalent positions of quadrupole nucleus in the crystal structure, (ii) by the local geometry (i.e. symmetry) of quadrupole nucleus surrounding, (iii) by the magnitude of nuclear spin $I$ and (iv) by the presence in nature of different isotopes of quadrupole nucleus. In particular, for copper there exist two naturally occurring isotopes $^{63}$Cu (69.2% natural abundance, $\gamma/2\pi=1.128$ kHz/Oe, $Q=-0.22$ barn) and $^{65}$Cu (30.8% natural abundance, $\gamma/2\pi=1.209$ kHz/Oe, $Q=-0.204$ barn) both having spin $I=3/2$ and thus two doubly degenerate $\pm 1/2$ and $\pm 3/2$ energy levels. Thus, for each isotope a transition between these levels yields a single NQR signal at a frequency:

$$^{63,65}\nu_Q = \frac{e \cdot {}^{63,65}QV_{ZZ}}{2h} \cdot \sqrt{1+\frac{1}{3}\eta^2} \ . \tag{4}$$

It should be noted that the NQR frequency $\nu_Q$, besides the quadrupole moment $eQ$ of the nucleus, depends on the certain arrangement of the surrounding ions through the parameters $V_{ZZ}$ and $\eta$. Since $\nu_Q$ is defined by two parameters in Eq. (4), $V_{ZZ}$ and $\eta$, it is impossible to obtain both parameters experimentally from the NQR spectra consisting of only one line for $I=3/2$. Usually it is done from the angular dependence of the NMR spectrum taken for a single crystal or from numerical simulations of NMR spectrum for the unaligned powder. In some cases, for determination of $\eta$ the 2D nutation NQR studies also could help. The values of $\eta$ at some temperatures for the triangular (plane) Cu(1) and tetrahedral Cu(2) sites in covellite CuS is given in Table II.

In a semi-empirical approach,[54] it is assumed that components of EFG tensor at Cu nuclei sites can be written as the sum of two terms – lattice and valence contributions:

$$V_{ZZ} = (1-\gamma_\infty) \cdot V_{latt} + (1-R_{val}) \cdot V_{val} \ . \tag{5}$$

The parameters $\gamma_\infty$, $R_{val}$ are the Sternheimer antishielding factors, which characterize the effects of charge density distortions induced by lattice electric field and non-fillness of orbitals of central atom. The first contribution arises from all ion charges outside the ion under consideration and can be calculated in a straightforward manner using the model of point charges (MPC):

$$V_{latt} = \sum_i \frac{q_i \cdot (3 \cdot \cos^2\theta_i -1)}{r_i^3} \ , \tag{6}$$

where $q_i$ and $r_i$ are the charge and the position of the $i$-th ion, respectively, $\theta_i$ – angle between the main axis of symmetry and the direction to the neighboring ion.

The second term in Eq. (5) arises from 3d and 4p unfilled shells and non-spherical distortions of inner orbitals of the subject Cu ion. Taking into account only holes in the Cu orbitals, the contributions of 3d and 4p shells can be written as

$$V_{val}^{(3d)} = -\frac{4}{7} \cdot e \cdot \langle r^{-3}\rangle_{3d} \cdot [N_{3d(3z^2-r^2)} - N_{3d(x^2-y^2)} - N_{3d(xy)} + \frac{1}{2}N_{3d(xz)} + \frac{1}{2}N_{3d(yz)}] , \tag{7.1}$$

$$V_{val}^{(4p)} = -\frac{4}{5} \cdot e \cdot \langle r^{-3}\rangle_{4p} \cdot [N_{4p(z)} - \frac{1}{2}N_{4p(x)} - \frac{1}{2}N_{4p(y)}] , \tag{7.2}$$

where $N_{3d(x,y,z)}$ and $N_{4p(x,y,z)}$ are the number of electronic holes in different 3d and 4p orbitals, the charge of an electron is given by $-e$.

The computation of total $V_{ZZ}$ constitutes a complex problem, since it requires a detailed knowledge of the structure and the population of the conduction band in metals. Thus, the comparison of $v_Q$ and $V_{ZZ}$, deduced from theoretical calculations on the basis of different structural models and approaches (see Eq. (5)), with experimental values (Eq. (4)) permits to determine the individual features of local electronic arrangement and peculiarities of chemical bonds[55] and, as consequence, NQR spectra can in general serve as phase-analytical diagnostics of different materials.

The longitudinal (spin-lattice) nuclear relaxation $T_1$ depends on specific sources of field fluctuations in crystal structures of compounds. Therefore the studies of the temperature dependences of nuclear relaxation often allow sensing the lattice dynamics and transport properties of the material.

## V. RESULTS

### V.1 Cu NQR spectra

It was found that copper NQR spectrum of synthetic stoichiometric sample of CuS (No.1) at 4.2 K consists of two copper doublets (i.e. the $^{63}$Cu and $^{65}$Cu isotopes lines): at 1.87 and 1.73 MHz ("low-frequency" doublet, Fig. 2) and at 14.88 and 13.77 MHz ("high-frequency" doublet, Fig. 3(a)). The $^{63}$Cu and $^{65}$Cu spectrum lines are identified based on the ratios of isotope quadrupole moments ($^{63}Q/^{65}Q = 1.081$) and their natural abundance ($^{65}A/^{63}A = 0.45$). The NQR spectrum of natural stoichiometric sample of CuS (No.5) is similar to that for sample No.1 and is not shown in paper. The presence of two spectral lines for both copper isotopes permits us to attribute NQR spectra to the two crystallographically non-equivalent sites of copper nuclei in CuS.

In contrast to stoichiometric CuS, copper NQR spectra of non-stoichiometric samples No.2-No.4 with chemical compositions $Cu_{1+x}S$ are more complex. As indicated in Fig. 3, at high-frequency range there are five copper NQR doublets, one of them corresponds to stoichiometric covellite CuS (exact coincidence of Cu line frequencies); the additional four Cu doublets belong to another structural phase. The NQR spectra, which pertain to this phase (Fig. 3), coincide in all non-stoichiometric samples and include the following lines: 16.95, 16.20, 15.25, 11.95 MHz for $^{63}$Cu and 15.67, 14.98, 14.11, 11.06 MHz for $^{65}$Cu. It should be noted that these eight Cu NQR lines have been observed earlier on the polycrystalline copper sulfide $Cu_{1.6}S$, also known as geerite.[56] Actually, in addition to covellite CuS, the existence of another phase in non-stoichiometric samples No.2-No.4 was proved by XRD analysis (Section III). At low-frequency range we also observed the multiplet NQR spectra with a few strongly overlapping Cu doublets (not shown here). The comparison of Cu NQR line positions allows us to conclude that two of them belong to the low-frequency doublet of stoichiometric CuS (Fig. 2).

Thus, the non-stoichiometric sulfides $Cu_{1+x}S$ with nonzero $x$ values appear to be multiphase and consist of a solid-state mixture of stoichiometric covellite CuS and geerite $Cu_{1.6}S$. It is interesting that the NQR signal intensities of $Cu_{1.6}S$ phase increase with increasing amount of additional copper ($x$) and, simultaneously, the signals intensities of CuS decrease according to the same proportion (Fig. 3). In the following, we will concentrate only on the temperature dependences of $^{63}$Cu isotopes NQR frequency, line-width and nuclear spin-lattice relaxation for covellite CuS phase (Figs. 2 and 3(a)). The Cu NQR spectra and nuclear relaxation of $Cu_{1.6}S$ will be published elsewhere.

### V.2 The temperature dependence of Cu NQR frequency and line-width

The dependence of the high-frequency $^{63}$Cu NQR line position on temperature is shown in Fig. 4(a). In general, the quadrupole frequency $v_Q$ decreases with increasing temperature without any significant anomalies. However, we focus here on two weak effects: the change of the slope in the $v_Q$ versus $T$ dependence at 65 K (i.e. near $T_{PT}$) and at 210 K. We mention here that the same effect at about 65 K is clearly seen in earlier studies.[27,28,57] In order to determine the approximate behavior of quadrupole frequency in the region 65–290 K we applied the following equation:[54]

$$\nu_Q(T) = \nu_Q(0) \cdot (1 - a \cdot T^b),  \quad (8)$$

where $\nu_Q(0)$, $a$ and $b$ are the fitting parameters. The best result of the fit, depicted in Fig. 4(a) by solid curve was obtained for $\nu_Q(0) = (14.95 \pm 0.02)$ MHz, $a = (1.7 \pm 0.5)*10^{-4}$ MHz/K and $b = (0.98 \pm 0.05)$. On the other hand, the $\nu_Q(T)$ dependence in the range 65–290 K could be divided into two regions, both of which can be well described by Eq. (8) with $b = 1$, i.e. linear function. The results of these fits, shown in inset of Fig. 4(a), were obtained for $\nu_Q(0) = (14.96 \pm 0.01)$ MHz, $a = (1.61 \pm 0.02)*10^{-4}$ MHz/K (in the range 65–210 K) and $\nu_Q(0) = (14.90 \pm 0.01)$ MHz, $a = (1.42 \pm 0.02)*10^{-4}$ MHz/K (in the range 210–290 K).

The Cu NQR line-shape was well fitted by Lorentzian function at all temperatures studied and its line-width was taken as full width at half maximum (FWHM). The temperature dependence of $\Delta\nu_Q$ is displayed in Fig. 4(b). As one can see, the NQR line-width increases linearly with decreasing $T$ in the range 290–55 K, but broadens more strongly below 55 K.

It is noteworthy that, compared with synthetic CuS (No.1), the natural CuS (No.5) demonstrates identical $T$-dependence of quadrupole frequency $\nu_Q$ (hence not shown), however NQR line-width $\Delta\nu_Q$ is broader by an additive constant value of about 15-20 kHz in the whole $T$-region, which is caused, obviously, by lattice defects in natural CuS with respect to synthetic analogue (Fig. 4(b)).

Due to the technical limitation, our NQR spectrometer cannot be used at frequencies below about 1.5 MHz. Since the $\nu_Q$ of low-frequency doublet decreases at higher temperatures, the $T$-dependences of $^{63}$Cu $\nu_Q$ and $\Delta\nu_Q$ have been studied only up to 30 K. Within experimental accuracy the values of $\nu_Q$ and $\Delta\nu_Q$ of low-frequency $^{63}$Cu NQR signal changes respectively from 1.87 MHz and 150 kHz at 4.2 K down to 1.78 MHz and 120 kHz at 30 K.

The $^{65}$Cu NQR line is about 1.1 times narrower than that of $^{63}$Cu for both position of copper at all studied $T$-regions; this states the quadrupolar mechanism of Cu NQR lines broadening.

*V.3 The temperature dependence of nuclear spin-lattice relaxation rates*

The longitudinal magnetization recovery curves for CuS in samples studied are well fitted to a single exponential function:

$$(M(\infty) - M(\Delta t))/M(\infty) = \exp(-(\Delta t / T_1)). \quad (9)$$

The dependences of $^{63}$Cu nuclear spin-lattice relaxation rates $1/^{63}T_1$ and $1/^{63}T_1T$ on temperature for both positions of copper in sample No.1 are presented in Fig. 5.

The relaxation rate $1/^{63}T_1$ of Cu nuclei, corresponding to high-frequency NQR line, linearly increases with increasing $T$ (Fig. 5(a)). However, more detailed measurements at low-$T$ revealed the change of slope in the $1/^{63}T_1$ versus $T$-dependence at about 17 K and 8 K (inset in Fig. 5(a)). As it should be seen in Fig. 5(b), the $1/^{63}T_1T$ exhibits $T$-independent behavior between 290 K and $T_{PT} = 55$ K, but below $T_{PT}$ demonstrates the strong falling down with the minimum at 8–9 K with the change of slope at 17 K. In contrast, in the region 7–1.47 K the $1/^{63}T_1T$ shows the abrupt increasing and, additionally, near 4 K the bend of $T$-dependence of $^{63}T_1T$ is observed (inset in Fig. 5(b)).

Remarkably, the values of $1/^{63}T_1$ and $1/^{63}T_1T$ for Cu nuclei, corresponding to low-frequency NQR line, show very similar $T$-dependence to that of "high-frequency" Cu (Fig. 5). Nevertheless, there are two differences. First, it is clearly seen that in studied $T$-region the low-frequency $^{63}$Cu nuclei are relaxing about 1.2 times faster than those of high-frequency $^{63}$Cu (inset in Fig. 5(a)). Second, the minimum in $1/^{63}T_1T$ occurs at higher temperature – near 16 K.

The isotopic ratio of $T_1^{-1}(^{65}Cu)/T_1^{-1}(^{63}Cu)$ can be used to identify the nature of relaxation process.[58] If the ratio is close to the squared ratio of gyromagnetic ratios $[\gamma(^{65}Cu)/\gamma(^{63}Cu)]^2 = 1.148$, the relaxation is caused by fluctuations of the local magnetic field. On the contrary, if the isotopic ratio is close to the ratio of squared nuclear electric quadrupole moments $[Q(^{65}Cu)/Q(^{63}Cu)]^2 = 0.856$, the resonance nucleus is feeling the fluctuations of the EFG tensor. In case of CuS, we have found that at all $T$-region studied the value of $T_1^{-1}(^{65}\text{Cu})$ nuclei is about 1.1 times larger than that of $^{63}$Cu for both positions of copper, i.e. relaxation process is magnetic in origin.

## VI. DISCUSSIONS

### *VI.1 NQR frequencies*

As already mentioned, the copper NQR spectrum of CuS is typical for two crystallographically non-equivalent positions of copper in the structure (Figs. 2 and 3(a)). Actually, covellite CuS is constructed from two different Cu complexes: triangular [Cu(1)-S(1)$_3$] and tetrahedral [Cu(2)-S(1)$_1$S(2)$_3$] (Fig. 1). The analysis of NMR spectra, independently attained by Itoh[27] and Saito[15], have shown that high-frequency satellite lines ($v_Q$ was estimated to be ~ 14.7 MHz) are assigned to three-coordinated Cu(1). Respectively, the low-frequency satellite lines ($v_Q$ ~ 1.5 MHz), are attributed to four-coordinated Cu(2).[15] Thus, the experimentally observed $^{63}$Cu NQR signals at 1.87 MHz (Fig. 2) and 14.88 MHz (Fig. 3(a) and Refs. 26-28) prove these results.

On the other hand, let us point out the studies of Abdullin and co-workers,[59] who experimentally revealed the dependence of the value of the $^{63}$Cu nuclei quadrupole frequency $v_Q$ upon the geometry of the most typical cases of Cu$^{1+}$ coordination environment in series of copper sulfides. In particular, in compounds with triangular complexes CuS$_3$ the value $v_Q$ lies within the range 20-23 MHz, whereas for tetrahedral coordination CuS$_4$, the $v_Q$ either equals zero (regular tetrahedron) or is small (distorted tetrahedron), usually less than 3 MHz. The theoretical calculations of EFG on Cu sites in CuS$_3$ units, carried out by Mulliken-Wolfsberg-Helmholtz (MWH) technique in the frame of MO-LCAO method, demonstrate the good agreement of computed values with experimental data, ternary sulfide Cu$_3$BiS$_3$ (wittichenite) being example of this.[60] It was shown that EFG (see Eq. (5)) is mainly formed by lattice term (90 %), and, at less extent, by Cu-3d and Cu-4p orbitals (altogether 10 %). The quite narrow range of the change of $v_Q$ in these sulfides can be explained by non-significant variations of Cu-S distances, S-Cu-S angles and polarity of chemical bonds.

In this sense, the shift of NQR frequency $v_Q$ of triangular (plane) copper Cu(1) in CuS towards lower values by about 5 MHz appears rather essential and unusual. This result indicates that the valence contribution is relevant for the EFG at the Cu(1) site. We suggest that the origin of $v_Q$ lowering is correlated to anomalously short Cu(1)-S distances in CuS compared to those in other sulfides (see Section II) and somehow related to the charge transfer between Cu-4p,3d and S-3p orbitals. Some qualitative considerations in favor of this are following. In principle, due to Bayer's influence of thermal-induced lattice vibrations, NQR frequency $v_Q$ must increase with decreasing distances as it takes place in most compounds.[55] However, in a number of cases the reducing of metal–ligand distances leads to the stronger hybridization of chemical bonds with subsequent charge transfer and its redistribution on the different orbitals, that can decrease the total EFG and $v_Q$ values.[55] The delafossite-based copper oxides CuMO$_2$ (M = Fe, Al, Ga) may serve as an example for such behavior.[61] It was illustrated that the reduction of only one Cu-O distance in linear CuO$_2$ units from 2.00 Å (CuAlO$_2$) down to 1.84 Å (CuGaO$_2$) exhibits the decrease of electronic density in Cu-4p$_z$ orbital and increase of the population of Cu-4p$_x$ and Cu-4p$_y$ orbitals, which play the key role in the experimentally observed lowering of total EFG value at the Cu site by about 1.5 MHz (see Eq. (7.2)). Since the O and S atoms have identical electronic configurations of outer shells (2s$^2$2p$^4$ and 3s$^2$3p$^4$, respectively), it is logical to suppose that the somewhat similar mechanism of $v_Q$ lowering takes place in CuS. It is not excluded that in the same manner the $v_Q$ lowering can be also caused by the charge redistribution in Cu-3d orbitals (see Eq. (7.1)). The influence of charge transfer on relaxation and Cu valence is discussed below (Section VI.4).

The significant deformations of CuS$_4$ tetrahedron break the cubic symmetry and, as it was mentioned, EFG on copper nucleus in this position becomes nonzero. Evidently, this case is realized for the tetrahedral Cu(2) sites in CuS (Fig. 2). Therefore our NQR spectra prove the crystallographic data[24] concerning the occurrence of low-symmetry distortions around Cu(2) sites below $T_{PT}$ (Fig. 1(b), Table I). To our knowledge, it is the first independent confirmation of the low-$T$ model of CuS structure.

The laborious search for a possible low-frequency NQR signal at $T$=77 K (above $T_{PT}$) found no any traces, which indicates that, if this signal exists, the value of $\nu_Q$ would be less than 1.5 MHz (technical limit of our NQR spectrometer; Section V.2). Actually, the estimation of $^{Cu(2)}\nu_Q$ by MPC calculations (Eq. (6)) for room-$T$ (Fig. 1(a), Table I) predicts the values of about 0.4 MHz, that signifies the negligible distortions of CuS$_4$ at $T$>55 K.

*VI.2 Knight shift*

Before analyzing the nuclear spin-lattice relaxation behavior in our NQR studies, it is expedient to comment recent unusual NMR data in CuS. The Cu NMR studies have shown that Knight shift $K$ for the Cu(1) has significant negative value (-1.4 % at 15 K)[15] but the explanation concerning the origin of such anomalous shift was not given. For transition metals the total Knight shift can be expressed through the respective hyperfine fields and susceptibilities as $K=K_s+K_{orb}+K_d(T)=\text{const}\cdot[F^s\cdot\chi_s+F^{orb}\cdot\chi_{orb}+F^d\cdot\chi_d]$, where only $K_d(T)$ depends on $T$.[58,62,63] Other contributions to total $K$ are considered to be negligible.[62] The first term $K_s$ arises from the contact interaction between the nuclear magnetic moment and spin-carrying electrons in the s-band, which create the hyperfine field $F^s$, and reflects the s-character Fermi level density of states (DOS). The second term is the orbital Knight shift originating from the orbital motion of electronic charges. The third term $K_d$ is created due to the exchange interaction between s electrons and the unpaired d electrons (referred to as "core polarization" effect) and related to the hybridization with d electrons.[58,64] Although $K_{orb}$ and $K_d$ are usually much smaller than $K_s$ in metals, they become competitive when the s-character Fermi-level DOS is significantly small. On the other hand, the exchange polarization of s states by d states generally contributes to a negative total shift $K$ due to the negative hyperfine field of d electrons,[58,62] as it takes place, for instance, in ternary carbides M$_2$AlC (M=Ti,V,Cr).[64] Evidently, $K_d$ overcomes the contribution from positive $K_s$ and $K_{orb}$ in CuS, leading to a negative total shift. Thus, the observed $K$[15] clearly reveals the enhancement of p-d hybridization between Cu-3d and S-3p orbitals. Such a result has been found to be consistent with that expected from the theoretical calculations,[35] indicating that p-d hybridization is substantial and that the ionic models (Cu$^{1+}$)$_3$(S$_2^{-2}$)(S$^{-1}$) or (Cu$^{1+}$)$_3$(S$_2^{-1}$)(S$^{-2}$) appears to be rather oversimplified.

Moreover, the total shift $K$ in CuS is larger at higher $T$ (-0.57 % at 60 K against -1.4 % at 15 K)[15] and, accordingly, the term $|K_d|$ should be smaller for Cu(1) above $T_{PT}$=55 K. This indicates that the extent of p-d hybridization between Cu(1)-3d and S(1)-3p orbitals with $T$-decreasing might be more significant. This conclusion confirms the predictions of theoretical calculations, according to which the 8 K structure is more stable than the room-$T$ modification.[25]

It is notable that the total shift $K$ for Cu(2) in CuS is positive and varies in the range 0.04-0.16 %.[15] These values are also small compared, for instance, to those in another transition metal - pure copper, in which the total Knight shift $K$ is equal to 0.23 %. This difference implies that $K_d$ for Cu(2) is also significant and comparable with $K_s+K_{orb}$, resulting in a small positive total shift.

*VI.3 Copper nuclear spin-lattice relaxation and Fermi level density of states*

The conduction electrons are known to govern the nuclear spin-lattice relaxation in metallic crystals.[58,62,63] The $T$-dependence of the relaxation rate $T_1^{-1}$ in this case is determined by fluctuations of the hyperfine magnetic field or the EFG with the frequencies ~10$^{15}$ sec$^{-1}$ created by the conducting spin-carrying charges at the nucleus site.[65,66] The isotopic ratio of Cu relaxation rates emphasizes the magnetic character of fluctuations (Section V.3). In the approach of free isotropic (3D) electronic gas and under the assumption that fluctuating magnetic fields are produced by the contact interaction of nucleus with the s-band electrons at the Fermi level,[62] the relaxation rate $T_1^{-1}$ can be expressed as:

$$\frac{1}{T_1} = \frac{8}{9}\pi^3 h^3 \gamma_e^2 \gamma_n^2 k_B \cdot T \cdot [N(E_F) \cdot \langle |\Psi(0)|^2 \rangle_F]^2, \qquad (10)$$

where $h$, $k_B$, and $T$ are the Planck constant, Boltzmann constant, and absolute temperature, respectively; $\gamma_n$ and $\gamma_e$ are the gyromagnetic ratios for Cu nuclei and electron, $\langle |\Psi(0)|^2 \rangle_F$ is the electronic spin density at the nucleus, and $N(E_F)$ represents the DOS of s-band electrons at the Fermi level, $N_s(E_F)$. Since all parameters in Eq. (10) are independent of $T$, a lot of metals exhibit the well-known constant $T_1 \cdot T$ behavior (also referred to as Korringa law). Actually, the constant $T_1 \cdot T$ holds approximately above $T_{PT}=55$ K (Fig. 5). Strictly speaking, if the core polarization effect occurs to be significant in transition metals (as it takes place for CuS, NMR of Cu(1) in which has negative Knight shift), it is quite appropriate to associate the observed $T_1$ with the effective value of $N(E_F)=N_s(E_F)+N_d(E_F)$, where $N_d(E_F)$ is d type of contribution to DOS.[62] The mixture of p state is treated only as part of the effective s type of DOS.[62]

At about $T_{PT}=55$ K, the relaxation rate demonstrates a surprising crossover from Korringa relation to unusual $T$-dependent $T_1 \cdot T$ behavior (Fig. 5), in spite of the fact that CuS appears to be a metal down to $T_c \approx 1.6$ K. Let us summarize the features that characterize the low-$T$ electronic behavior.

First, we adduce following arguments in favor of the assumption that low-$T$ relaxation is intrinsic in nature, i.e., that it is inherent in the compound itself. We have checked the $T$-dependences of $1/^{63}T_1$ and $1/^{63}T_1T$ in samples No.2–No.5 and found that data are identical to those in sample No.1. Furthermore, earlier studies of nuclear relaxation for Cu(1) in CuS[27] also supply data for deviations from linearity in $1/^{63}T_1T$ below $T_{PT}$, but without attention paid to such behavior. The identical results for different CuS samples indicate that the observed low-$T$ peculiarities are not associated with any magnetic impurities or lattice defects.

Second, although we do not have conclusive evidence concerning the origin of the unusual $T_1 \cdot T$ behavior, the following observation suggests that collective electronic charges coupled motion below $T_{PT}$ is important. Both non-equivalent sites of Cu exhibit a very similar dependence of relaxation on $T$, which gives the constant ratio of the two relaxation times $^{Cu(1)}T_1/^{Cu(2)}T_1 \approx 1.2$ (Fig. 5). Since, in any case, the nuclear relaxation is governed by conduction charges, our result implies that *common* electronic dynamics must exist for the two *non-equivalent* Cu(1) and Cu(2), i.e. they "feel" not the individual, but the *same* conduction band. Therefore, taking into account the 2D character of conductivity in CuS below $T_{PT}$ in the direction perpendicular to the $c$-axis (Section II), we can suggest that the electronic charges are delocalized not only in the plane of Cu(1)-S(1)$_3$ units as "the two-dimensional sea",[35] but also between two sites of copper - Cu(1) and Cu(2) through the bridging S(1) ion, creating in such a way the "waves" in this sea (Fig. 1).

Third, it is notable that similar deviations of $1/^{63}T_1T$ from linearity at $T>T_c$ are well-known features for most layered HTSC, in particular, YBa$_2$Cu$_3$O$_{6+x}$.[67,68] Since in usual metals $1/^{63}T_1T \sim [N(E_F)]^2$ (see Eq. (10)), the appreciable decreasing of $1/^{63}T_1T$ is widely explained by decreasing of $N(E_F)$. This effect is often referred to as opening of "gap" or "pseudogap" in the normal-state DOS. The appearance of such pseudogap in HTSC is also reflected in $T$-dependences of Knight shift, heat capacity, magnetic susceptibility, electrical resistivity, neutron scattering, angle resolved photoemission spectroscopy (ARPES) and scanning tunneling microscope (STM) studies.[67,68] It looks like similar deviation of $1/^{63}T_1T$ from constant value in CuS (Fig. 5) points to the occurrence of some gap in this material below $T_{PT}$. Actually, we remind that different anomalies in CuS below $T_{PT}$ have been also found in heat capacity, electrical resistivity and neutron diffraction investigations (Section II). However, intriguing Knight shift studies were made up without desirable experimental accuracy, susceptibility data are inconsistent and ARPES, STM investigations of CuS are lacking. Therefore, for examination of our suggestion it would be expedient to perform the respective detailed studies. We also add here that interpretations of pseudogap state nature in HTSC and its influence on the superconductivity are manifold,[68] including among others a charge-density wave (CDW) model.[69] According to this approach, the pseudogap transformations of electronic spectra take place due to strong scattering of electrons on CDW.

Fourth, the upturn of $1/^{63}T_1T$ for both Cu(1) and Cu(2) below 8 K and 16 K, respectively, suggests the presence of gapless mode. Its origin is not clear for us, but similar behavior of relaxation in some HTSC is also observed.[70]

Fifth, it should be noted that $1/^{63}T_1T$ for Cu(1) demonstrates the bends at about 4 K and 17 K (Fig. 5), which, most likely, signify the occurrence of some changes in internal dynamics in CuS. Interestingly, these points become apparent also in heat capacity[37] and ac magnetic susceptibility[11] measurements.

### VI.4 Copper valence and charge-density waves

The important piece of information that can be extracted from the NQR spectra data is the absence of magnetic ordering in CuS down to low $T$. Since $1 \cdot \mu_B$ of spin moment in 3d magnetically ordered state typically gives rise to the strong static field, a sharp single line for each NQR transition of Cu (see Section V.1) indicates that no internal static field due to any type of magnetic ordering is present at the Cu sites in CuS for the temperature range studied. Such a field would magnetically broaden, split or shift the NQR lines as it does for the Cu signal in antiferromagnets $YBa_2Cu_3O_{6+x}$[71] or transition metal sulfide $CuFe_2S_3$ (cubanite).[72] These data are also supported by quadruple character of the line broadening for both $^{63}$Cu NQR signals (Section V.2). Our conclusion is in a good consistent with EPR studies of CuS, showing no paramagnetic divalent $Cu^{2+}$ signal at room-$T$ and low-$T$,[23] that can suggests the mixed-valence state of Cu besides $Cu^{1+}$.[23]

Actually, the observation of $^{63}$Cu NQR lines cannot exclude the presence of so-called "exchange-narrowed" paramagnetic Cu ions,[73] in which the total number of 3d electrons, $n_d$, changes between values 9.0 (corresponds to paramagnetic $Cu^{2+}$, electronic spin $S=1/2$) and 10.0 (diamagnetic $Cu^{1+}$, electronic spin $S=0$),[74] the NQR spectra of binary compounds $CuF_2$, $CuBr_2$ and $CuCl_2$ being example of this.[73] In fact, on the basis of Cu-2p XPS studies of CuS, it was estimated that the $n_d$ value can differ from 10.0 but it is more than 9.5,[74] i.e. Cu is closed to be monovalent $Cu^{1+}$.

We turn to the negative shift $K_d$ for Cu(1) (see Section VI.2): this contribution is proportional to the number of *unpaired* 3d electrons.[58,62] Indeed, the presence of unpaired d electrons was proposed earlier on the base of XPS spectrum analysis.[75] This strongly advocates that Cu is not monovalent, since in case of $Cu^{1+}$ all 3d electrons are paired ($n_d=10$). Therefore we can suggest that the valence of copper in CuS has non-integer value and intermediates in the range $Cu^{1+}$ and $Cu^{1.5+}$ (i.e. $9.5<n_d<10.0$). This is consistent with the interpretation of some crystal-chemical features of CuS,[21] according to which the valence of Cu(1) and Cu(2) should satisfy the value of $Cu^{1.3+}$. It is interesting to note that some recent studies reveal the "d count" in the Cu sulfides to be intermediate between $3d^9$ and $3d^{10}$, although these sulfides were initially classified as nominally monovalent or divalent Cu compounds (for example, chalcopyrite $CuFeS_2$[74] and tennantite $Cu_{12}As_4S_{13}$[76,77]). Probably, this tendency is the consequence of some "antipathy of Cu for a $3d^9$ configuration and the stability of Cu $3d^{10}$ in Cu sulfides".[48]

Let us now consider the dynamic effects. Theoretical calculations of CuS conduction band predict that "electronic charge should flow from the $4t_2$-orbital on the tetrahedral Cu(2) to the $4e$-orbital on the triangular Cu(1)" through the bridging S(1).[78] Our relaxation studies strongly support this suggestion (Section VI.3). Moreover, it is proposed that because of fast charge mobility this electronic transfer can be realized as Cu valence fluctuation.[45]

In fact, it has been argued that the $T$-dependence of NQR frequency $^{Cu(1)}\nu_Q$ in CuS can be understood in terms of charge fluctuations in Cu(1)-S(1)-Cu(2) bonds. It is known that in most non-cubic metals $\nu_Q(T)$ can be well reproduced by the empirical Eq. (8) with $b=1.5$; this relation is often referred to as "$T^{3/2}$ law".[54] Generally today, thermal vibrations of the host lattice atoms are regarded as mainly responsible for such simple and universal relation. However, CuS shows the change of slope in the $\nu_Q(T)$ near 210 K, which exhibits another relation – parameter $b$ in Eq. (8) is close to be

1.0 (notably, Ref. 24 found that the diffraction reflections, which become split below $T_{PT}$, are already broadened at some 150–200 K). To our opinion, the $v_Q(T)$ dependence in the range 65–290 K can be described more exactly by two linear functions (Section V.2). But in any case $^{Cu(1)}v_Q(T)$ in CuS do not follows to $T^{3/2}$ law. The very similar $v_Q(T)$ behavior was found in mixed-valence metal EuCu$_2$Si$_2$.[79] It was shown that emitting of conduction electron by neighboring Eu ion causes the fluctuation between two Eu electronic configurations. Such valence instability influences the Cu quadrupole interactions ($V_{ZZ}$) and, as a consequence, the $T^{3/2}$ law becomes not valid. Since in the range 60–290 K there are no any structural changes in CuS[24] and $\eta \approx 0$ (Table II), the $T$-dependence of $^{Cu(1)}v_Q$ is determined only by $V_{ZZ}$. Therefore, by analogy, we suggest that bridging S(1) ion can provide minor charge transfer between Cu(1) and Cu(2) in some fluctuating regime. The strong hybridization of Cu(1) and Cu(2) conduction bands, as it is seen from low as compared to other sulfides Cu(1) NQR frequency (Section VI.1) and relaxation measurements (Section VI.3), should allow this transfer.

It is exciting that our NQR measurements on "plane" Cu(1) in CuS strongly resemble the NQR studies, performed on Cu plane sites in HTSC YBa$_2$Cu$_3$O$_7$.[80] Sharp increase of NQR line-width and falling down of $1/^{63}T_1T$ led the authors to the conclusion that *charge modulation in the form of CDW* takes place in this material. In general, the CDW comes from the periodic redistribution of electronic charges due to small ionic movement near their equilibrium position in the crystal lattice. In the case of quasi-2D (layered) materials CDW are formed below some critical temperature and manifest themselves in the appearance of energy gap in electronic spectra on overlapping Fermi surface patches (i.e. to partial loss of metallic properties) and DOS modulations. Actually, triangular coordinated Cu(l) ions have a large anisotropic thermal motion at room-$T$ and strange oscillations of cross-sections of the Fermi surface were detected.[25] Thus, the similarity of NQR data allow us to suppose that CDW can exist in CuS and that CDW manifest themselves as divergence of $1/^{63}T_1T$ from constant value below $T_{PT}$ and unusual $v_Q(T)$ dependence.

The "pure" NQR studies of layered metal dichalcogenides are relatively rare and it is difficult to compare our NQR results in CuS with those in other dichalcogenides. It is known that low-$T$ superconductor NbSe$_2$, in which the presence of CDW is proved,[17] also demonstrates unusual $^{Nb}v_Q(T)$ dependence.[81]

*VI.5 Phase transition in CuS*

The comparison of our $^{63}$Cu NQR spectrum for CuS (Figs. 2 and 3(a)) and $^{63}$Cu NQR spectrum for selenium analog of CuS – α-CuSe (klockmannite)[82] can be used for discussions of reasons for PT in CuS at 55 K. It was found that α-CuSe is characterized by the 13 lines with NQR frequencies in the range of 12.7–2.09 K at 4.2 K. This points to the presence of strong deformations of the crystal (compared to CuS) and produces evidence that, strictly speaking, CuS and α-CuSe are *not isostructural* as it was supposed earlier.[83]

To this moment, it is clearly seen that insertion of Se atoms instead of S in CuS results in *conversion* of Cu(1) threefold units into distorted fourfold (since $^{63}$Cu NQR frequencies are rather low) and that α-CuSe has 13 non-equivalent Cu positions in at the least. To our opinion, the initial occupancy of Se at the S(2) sites[38] leads to the approaching of some Se(2) atoms to Cu(1) sites and, as consequence, to the formation of "new" distorted tetrahedrons [Cu(1)-Se(1)$_3$Se(2)$_1$] in CuSe instead of "old" triangular units [Cu(1)-S(1)$_3$] in CuS (Fig. 1). In this case such deformations would promote the approaching of Cu(1) and Cu(2) ions to each other and creation of effective interaction between them, as it was proposed for CuS[24] and α-CuSe.[84] Upon cooling this interaction can stimulate the hexagonal-to-orthorhombic transition, as for CuS[24] and α-CuSe.[84] Such scenario is supported by the dependence of $T_{PT}$ value on Se amount in mixed samples CuS$_{1-x}$Se$_x$ (0≤x≤1).[38] Actually, due to longer Cu(1)-Cu(2) bonds in CuS than in CuS$_{1-x}$Se$_x$, this PT for CuS occurs at lower $T$. Interestingly, the formation of effective metal-metal interactions in binary sulfides often causes the structural transformation as, for instance, in VS$_3$ (patronite).[2]

# VII. CONCLUSIONS

We have studied the electronic behavior of 2D transition metal CuS (covellite) using NQR as a probe in a wide temperature range. We have found that CuS exhibits not one, but two $^{63}$Cu NQR signals at 1.87 and 14.88 MHz (4.2 K), which are attributed to Cu(2) and Cu(1) nucleus, respectively. The "pure" NQR spectra in CuS are an experimental proof that no magnetic ordering occurs in this compound. The direct observation of low-frequency $^{63}$Cu NQR signal in CuS demonstrates the serious distortions of [Cu(2)-S$_4$] units at low temperatures. The high-frequency $^{63}$Cu NQR signal is placed out of frequency range 20–23 MHz typical for triangularly coordinated Cu with three S atoms in other copper sulfides and, therefore, this serves as indication of the strong hybridization of Cu(1)–S(1) bonds in [Cu(1)-S$_3$] units. The temperature dependence of Cu(1) quadrupole frequency $v_Q$, line-width $\Delta v_Q$ and nuclear spin-lattice relaxation $T_1$, which so far had never been investigated so precisely for CuS, altogether display the occurrence of structural phase transition at about 55 K. It has been argued that this transition is stimulated by Cu(1)-Cu(2) interactions. Unusual behavior of nuclear relaxation rates of Cu(1) and Cu(2) provide evidence that this phase transition is accompanied by electronic spectra transformations, which could be interpreted as the formation of the energy gap. Moreover, nuclear relaxation rates of both Cu ions point to the strong hybridization of Cu(1) and Cu(2) conduction bands through S(1) orbitals, leading to the presence of charge transfer in Cu(1)–S(1)–Cu(2) bonds even in low-temperature anisotropic 2D regime. The low-$T$ relaxation of $^{63}$Cu is also indicative of a significant contribution of d density of states at the Fermi level. Analysis of NQR spectra and literature data allows us to conclude that valence state of both Cu is not strictly monovalent Cu$^{1+}$ or divalent Cu$^{2+}$, but intermediate with dominant of former with average value ≈ 1.3+. In addition, the dependence of Cu(1) quadrupole frequency $v_Q$ on temperature, not typical for "simple" metals, have been interpreted from the viewpoint of Cu valence fluctuation in the vicinity of the average value. We have suggested that charge-density waves in the CuS electronic structure could be responsible for the appearance of the energy gap and may be connected to charge transfer and Cu valence instability. Nevertheless, the detailed microscopic picture of this phenomenon is unclear at present.

Finally, from our results on CuS it is clearly seen that this compound exhibits interesting properties, which have never been observed in previous investigations, and appears to be a good example of layered dichalcogenide class of materials. We believe that the reported NQR studies provides an advanced understanding of CuS electronic characteristics and can be considered as a base for further studies, either experimental or theoretical.

## Acknowledgments

Authors thank Prof. Eremin M.V. (Kazan State University) for constructive advices, expressed during discussion of the form and content of this paper. Authors thank Prof. Kal'chev V.P. (Kazan Armor Institute) for providing of his original Ph.D. thesis. Authors thank Prof. Krinari G.A. (Kazan State University) for XRD identification of samples. This work was partly supported by Russian Foundation for Basic Research under Grant 06-02-17197a.

## References

[1] D.J. Vaughan and J.R. Craig, *Mineral Chemistry of Metal Sulphides* (Cambridge University Press, Cambridge, England, 1978)

[2] A.S. Marfunin, *Physics of minerals and inorganic materials: an introduction* (Springer-Verlag, Berlin, 1979) [English translation of Russian original edition: Publishing House Nedra, Moscow, USSR, 1974]

[3] V.M. Izoitko, *Technological mineralogy and estimation of ore* (Publishing House Nauka, St. Petersburg, Russia, 1997)

[4] C.I. Pearce, R.A.D. Pattrick, and D.J. Vaughan, Rev. Mineral. Geochem. **61**, 127 (2006)

[5] See, for example, L.A. Isac, A. Duta, A. Kriza, I.A. Enesca, and M. Nanu, J. Phys.: Confer. Series **61**, 477 (2007)

[6] See, for example, A.J Aguiar, C.L.S. Lima, Y.P. Yadava, L.D.A Tellez, J.M. Ferreira, and E. Montarroyos, Physica C **341**, 593 (2000)

[7] J.-S. Chung and H.-J. Sohn, J. Power Sources **108**, 226 (2002)

[8] See, for example, W.U. Dittmer and F.C. Simmel, Appl. Phys. Let. **85**, 633 (2004)

[9] W. Meissner, Z. für Physik **58**, 570 (1929)

[10] W. Buckel and R. Hilsch, Z. für Physik **128**, 324 (1950)

[11] F. Di Benedetto, M. Borgheresi, A. Caneschi, G. Chastanet, C. Cipriani, D. Gatteschi, G. Pratesi, M. Romanelli, and R. Sessoli, Eur. J. Mineral **18**, 283 (2006)

[12] J.G. Bednorz and K. Muller, Z. für Physik B **64**, 189 (1986)

[13] See, for example, I. R. Mukhamedshin, H. Alloul, G. Collin, and N. Blanchard, Phys. Rev. Lett. **94**, 247602 (2005)

[14] H. Nozaki, K. Shibata, and N. Ohhashi, J. Solid State Chem. **91**, 306 (1991)

[15] S.-h. Saito, H. Kishi, K. Nie, H. Nakamaru, F. Wagatsuma, and T. Shinohara, Phys. Rev. B **55**, 14527 (1997)

[16] R. Könenkamp, Phys. Rev. B **38**, 3056 (1988)

[17] A.H. Castro Neto, Phys. Rev. Lett. **86**, 4382 (2001)

[18] A.A. Kordyuk, S.V. Borisenko, V.B. Zabolotnyy, R. Schuster, D.S. Inosov, R. Follath, A. Varykhalov, L. Patthey, and H. Berger, arXiv: 0801.2546 (2008)

[19] D.W. Shen, B.P. Xie, J.F. Zhao, L.X. Yang, L. Fang, J. Shi, R.H. He, D.H. Lu, H.H. Wen, and D. L. Feng, Phys. Rev. Lett. **99**, 216404 (2007)

[20] H. Barath, M. Kim, J.F. Karpus, S.L. Cooper, P. Abbamonte, E. Fradkin, E. Morosan, and R.J. Cava, Phys. Rev. Lett. **100**, 106402 (2008)

[21] H. T. Evans Jr. and J. A. Konnert, Am. Mineral. **61**, 996 (1976)

[22] J.A. Tossell, Phys. Chem. Minerals **2**, 225 (1978)

[23] K. Bente, Mineral. Petrol. **36**, 205 (1987)

[24] H. Fjellvåg, F. Grønvold, S. Stølen, A.F. Andresen, R. Müller-Käfer, and A. Simon, Z. für Kristallogr. **184**, 111 (1988)

[25] W. Liang and M.-H. Whangbo, Solid State Commun. **85**, 405 (1993)

[26] R.S. Abdullin, V.P. Kal'chev, and I.N. Pen'kov, Dokl. Akad. Nauk SSSR [Sov. Phys.–Dokl.] **294**, 1439 (1987)

[27] Y. Itoh, A. Hayashi, H. Yamagata, M. Matsumura, K. Koga, and Y. Ueda, J. Phys. Soc. Japan **65**, 1953 (1996)

[28] H. Tnabe, H. Kishi, Nakamaru, S.-h. Saito, F. Wagatsuma, and T. Shinohara, Meet. Abstr. Phys. Soc. Japan **52**, 710 (1997)

[29] L.G. Berry, Am. Mineral. **39**, 504 (1954), and reference therein of more earlier studies

30 A.S. Povarenykh, *Crystal chemical classification of minerals* (Plenum Press, New York, USA, 1972) [English translation of Russian original edition: Publishing House Naukova Dumka, Kiev, USSR, 1966]

31 N.V. Belov and E.A. Pobedimskaya, Kristallografiya **13**, 969 (1968) [English translation: Sov. Phys.–Crystallogr. **13**, 843 (1969)]

32 R. Kalbskopf, F. Pertlik, and J. Zemann, Tschermaks Miner. Petrogr. Mitt. **22**, 242 (1975)

33 M. Ohmasa, M. Suzuki, and Y. Takeuchi, Mineral. J. (Japan), **8**, 311 (1977)

34 A. Putnis, J. Grace, and W.E. Cameron, Contr. Mineral. Petrol. **60**, 209 (1977), and reference therein

35 H.J. Gotsis, A.C. Barnes, and P. Strange, J. Phys.: Condens. Matter **4**, 10461 (1992)

36 M. Isino and E. Kanda, J. Phys. Soc. Japan **35**, 1257 (1973)

37 E.F. Westrum Jr., S. Stølen, and F. Grønvold, J. Chem. Thermodynamics **19**, 1199 (1987)

38 H. Nozaki, K. Shibata, M. Ishii, and K. Yukino, J. Solid State Chem. **118**, 176 (1995)

39 M. Ishii, K. Shibata, and H. Nozaki, J. Solid State Chem. **105**, 504 (1993)

40 T. Nakajima, M. Isino, and E. Kanda, J. Phys. Soc. Japan **28**, 369 (1970)

41 T. Chattopadhyay and A. R. Chetal, J. Phys. Chem. Solids **46**, 427 (1985)

42 M. Grioni, J.B. Goedkoop, R. Schoorl, F.M.F. de Groot, J.C. Fuggle, F.Schäfers, E.E. Koch, G. Rossi, J.-M. Esteva, and R.C. Karnatak, Phys. Rev. B **39**, 1541 (1989)

43 I.B. Borovskii and I.A. Ovsyannikova, Zh. Eksp. Teor. Fiz. **37**, 1458 (1959) [English translation; Sov. Phys.–JETP **37**, 1033 (1959)]

44 L. Nakai, Y. Sugitani, K. Nagashima, and Y. Niwa, J. Inorg. Nucl. Chem. **40**, 789 (1978)

45 J.C.W. Folmer and F. Jellinek, J. Less-Common Met. **76**, 153 (1980)

46 D.L. Perry and J.A. Taylor, J. Mater. Sci. Lett. **5**, 384 (1986)

47 G. van der Laan, R.A.D. Pattrick, C. M. B. Henderson, and D. J. Vaughan, J. Phys. Chem. Solids **53**, 1185 (1992)

48 R.A.D. Pattrick, J.F.W. Mosselmans, J.M. Charnock, K.E.R. England, G.R. Helz, C.D. Garner, and D.J. Vaughn, Geochim. Cosmochim. Acta **61**, 2023 (1997)

49 S.W. Goh, A.N. Buckley, R.N. Lamb, R.A. Rosenberg, and D. Moran, Geochim. Cosmochim. Acta **70**, 2210 (2006)

50 E.Z. Kurmaev, J. van Ek, D.L. Ederer, L. Zhou, T.A. Callcott, R.C.C. Perera, V.M. Cherkashenko, S.N. Shamin, V.A. Trofimova, S. Bartkowski, M. Neumann, A. Fujimori, and V.P. Molosha, J. Phys.: Condens. Matter **10** 1687 (1998)

51 D. Li, G.M. Bancroft, M. Kasrai, M.E. Fleet, X.H. Feng, B.X. Yang, and K.H. Tan, Phys. Chem. Minerals **21**, 317 (1994)

52 C. Sugiura, H. Yamasaki, and S. Shoji, J. Phys. Soc. Japan **63**, 1172 (1994)

53 A.P. Bussandri and M.J. Zuriaga, J. Magn. Reson. **131**, 224 (1998)

54 E.N. Kaufmann and R.J. Vianden, Rev. Mod. Phys. **51**, 161 (1979)

55 G.K. Semin, T.A. Babushkina, and G.G. Yakobson, *Nuclear quadrupole resonance in chemistry* (John Wiley & Sons, New York, 1975) [English translation of Russian original edition: Publishing House Khimiya, Leningrad, USSR, 1972]

[56] V.N. Anashkin, T.A. Kalinina, V.L. Matukhin, I.N. Pen'kov, and I.A. Safin, Zapiski RMO [Proceedings RMS] **5**, 59 (1994)

[57] Itoh (Ref. 27) studied the $^{Cu(1)}v_Q(T)$ with small amount of steps within 100–300 K, whereas Tnabe (Ref. 28) measured the value of $^{Cu(1)}v_Q$ only up to 80 K. Therefore the bend of $v_Q(T)$ at 210 K in these studies is not found.

[58] J. Winter, *Magnetic Resonance in Metals* (Clarendon Press, Oxford, England, 1971)

[59] R.S. Abdullin, V.P. Kal'chev, and I.N. Pen'kov, Phys. Chem. Minerals **14**, 258 (1987); V.P. Kal'chev, Ph.D. thesis, Kazan State University, 1988.

[60] I.N. Pen'kov, R.S. Abdullin, N.B. Yunusov, and N.V. Togulev, Izv. Akad. Nauk SSSR Ser. Fiz. [Bull. Acad. Sci. USSR, Phys. Ser.] **42**, 2104 (1978)

[61] R.S. Abdullin, I.N. Pen'kov, and N.B. Yunusov, Izv. Akad. Nauk SSSR Ser. Fiz. [Bull. Acad. Sci. USSR, Phys. Ser.] **45**, 1787 (1981)

[62] A. Narath, in *Hyperfine Interactions*, edited by A.J. Freeman and R.B. Frenkel (Academic Press, New York, USA, 1967)

[63] C.P. Slichter, *Principles of Magnetic Resonance* (Springer-Verlag, Berlin, Germany, 1996)

[64] C.S. Lue, J.Y. Lin and B.X. Xie, Phys. Rev B **73**, 035125 (2006)

[65] A.H. Mitchell, J. Chem. Phys. **26**, 1714 (1957)

[66] A.R. Kessel, Fiz. Met. Metalloved. [Phys. Met. Metallogr. USSR] **23**, 837 (1967)

[67] N.M. Plakida, *High-temperature superconductivity* (Springer-Verlag, Berlin, Germany, 1995); D. Brinkmann, J. Mol. Struct. **345**, 167 (1995)

[68] M.V. Sadovskii, Phys. Usp. **44**, 515 (2001)

[69] A.V. Dooglav, M.V. Eremin, Yu.A. Sakhratov, and A.V. Savinkov, JETP Lett. **74**, 103 (2001); M.V. Eremin, I. Eremin, and A. Terzi, Phys. Rev. B **66**, 104524 (2002)

[70] K. Ishida, K. Yoshida, T. Mito, Y. Tokunaga, Y. Kitaoka, K. Asayama, Y. Nakayama, J. Shimoyama, and K. Kishio, Phys. Rev. B **58**, R5960 (1998)

[71] A.V. Dooglav, A.V. Egorov, I.R. Mukhamedshin, A.V. Savinkov, H. Alloul, J. Bobroff, W.A. MacFarlane, P. Mendels, G. Collin, N. Blanchard, P.G. Picard, P.J.C. King, and J. Lord, Phys. Rev. B **70**, 054506 (2004)

[72] R.S. Abdullin, V.P. Kal'chev, and I.N. Pen'kov, Fiz. Tverd. Tela (Leningrad) [Sov. Phys.–Solid State] **22**, 2862 (1980)

[73] T.J. Bastow, I.D. Campbell, and K.J. Whitfield, Solid State Commun. **33**, 399 (1980)

[74] C.I. Pearce, R.A.D. Pattrick, D.J. Vaughn, C.M.B. Henderson, and G. van der Laan, Geochim. Cosmochim. Acta **70**, 4635 (2006)

[75] T. Novakov, Phys. Rev. B **3**, 2693 (1971)

[76] R.R. Gainov, A.V. Dooglav, and I.N. Pen'kov, Solid State Commun. **140**, 544 (2006)

[77] R.R. Gainov, A.V. Dooglav, I.N. Pen'kov, I.R. Mukhamedshin, A.V. Savinkov, and N.N. Mozgova, Phys. Chem. Minerals **35**, 37 (2008)

[78] D.J. Vaughan and J.A. Tossell, Phys. Chem. Minerals **9**, 253 (1983)

[79] E.V. Sampathkumaran, L.C. Gupta, and R. Vijayaraghavan, Phys. Rev. Lett. **43**, 1189 (1979); E.V. Sampathkumaran, L.C. Gupta, and R. Vijayaraghavan, J. Mol. Struct. **58**, 89 (1980)

[80] B. Grévin, Y. Berthier, and G. Collin, Phys. Rev. Lett. **85**, 1310 (2000)

[81] D.R. Torgeson and F. Borsa, Phys. Rev. Lett. **37**, 956 (1976)

[82] V.P. Kal'chev, I.N. Pen'kov, and T.A. Kalinina, Zapiski VMO [Proceedings RMS] **6**, 81 (1989)

[83] R.D. Heyding and R.M. Murray, Can. J. Chem. **54**, 842 (1976)

[84] V. Milman, Acta Cryst. B **58**, 437 (2002), and reference therein

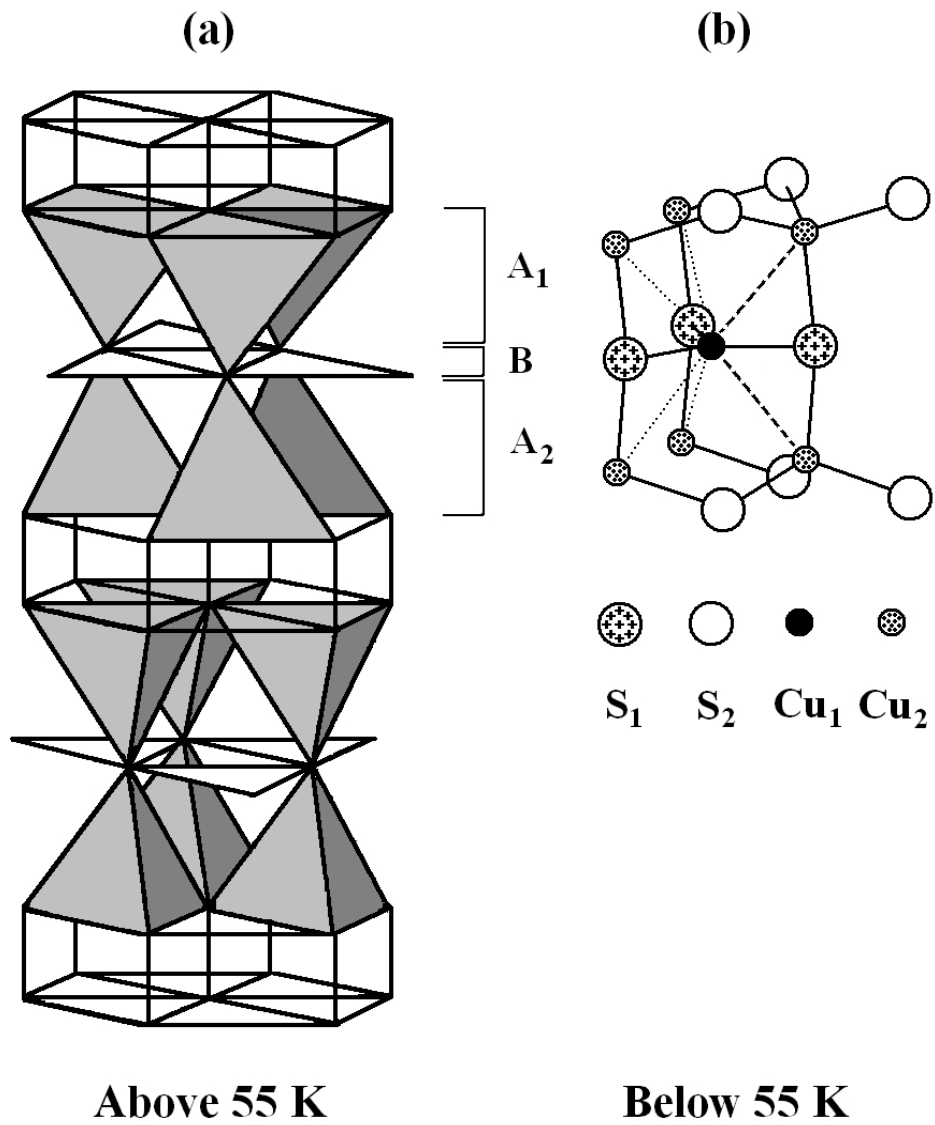

**FIG.1. (a)** Crystal structure of the covellite CuS above $T_{PT} = 55$ K (reproduced from Ref. 30). The main structural units are the CuS$_4$–tetrahedra (layers A$_1$ and A$_2$) and the CuS$_3$–triangles (layer B). **(b)** The fragment of crystal structure of CuS below $T_{PT} = 55$ K (reproduced from Ref. 24). It is clearly seen that the CuS$_4$–tetrahedra are distorted. For details, see Section II.

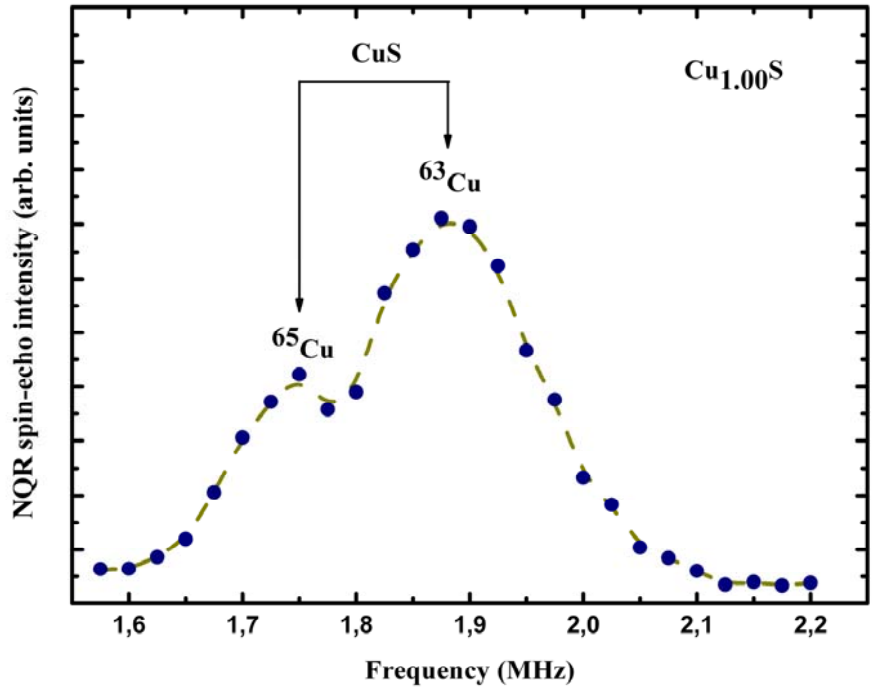

**FIG.2.** Low-frequency copper NQR spectrum for synthetic $Cu_{1.00}S$ (sample No.1) at 4.2 K (circles). Arrows point to the positions of $^{63,65}Cu(2)$ NQR signals of covellite CuS phase. Dashed curve is the fit to the data. For details, see Sections V.1 and VI.1.

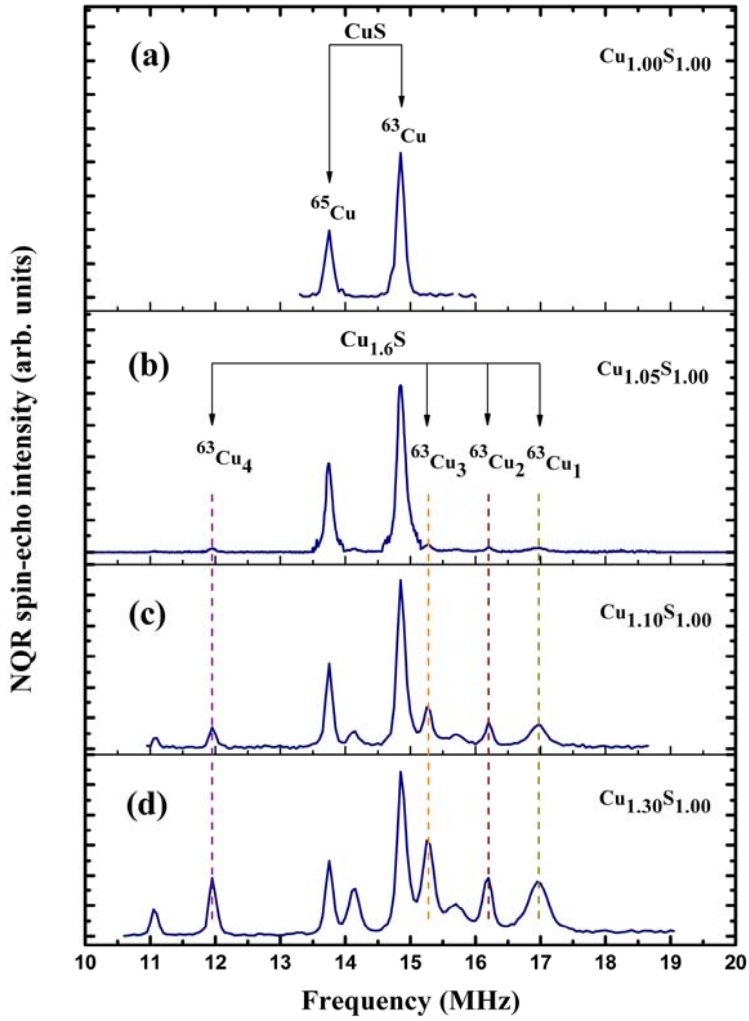

**FIG.3.** High-frequency copper NQR spectra for synthetic $Cu_{1+x}S$ for *x* values of 0, 0.05, 0.10 and 0.30 at 4.2 K (solid curves). Arrows point to the positions of $^{63,65}Cu(1)$ NQR signals of covellite CuS phase in samples No.1-No.4 (**(a)**–**(d)**, respectively). For details, see Sections V.1 and VI.1. The vertical dashed lines point to the positions of $^{63}Cu$ NQR signals of geerite $Cu_{1.6}S$ phase in samples No.2-No.4 (**(b)**–**(d)**, respectively). For details, see Section V.1.

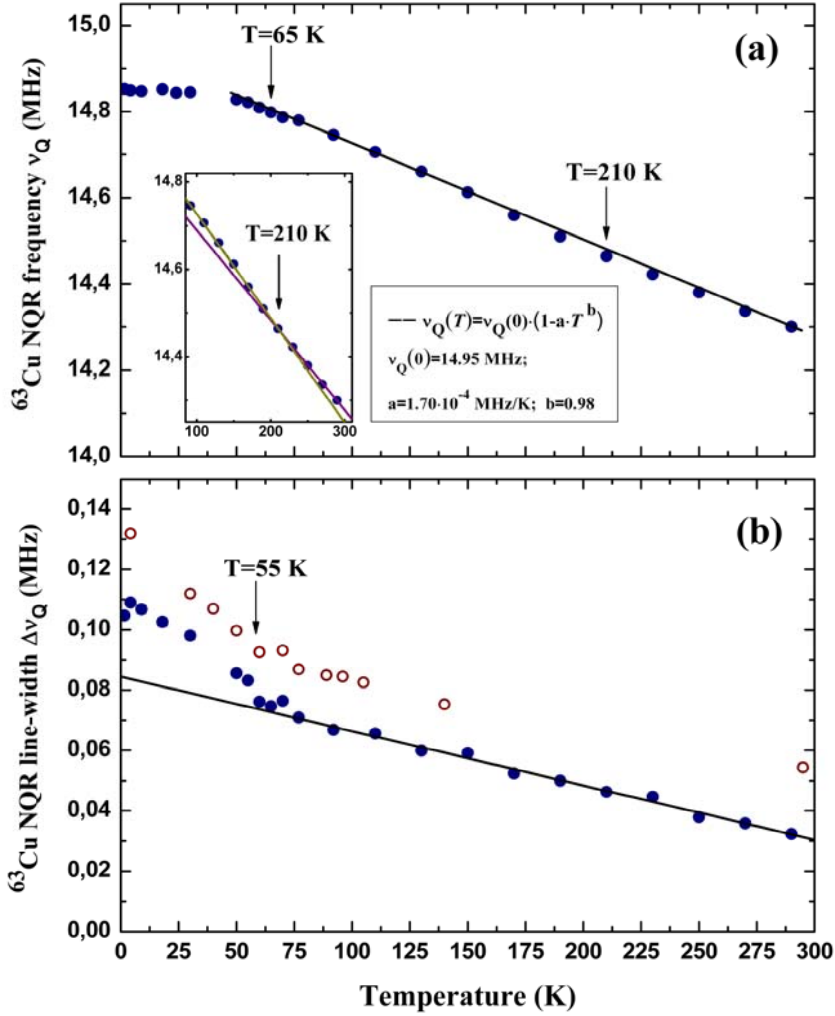

**FIG.4. (a)** The temperature dependence of $^{63}$Cu(1) NQR frequency $v_Q$ for CuS phase in sample No.1 (closed circles) with the fit of data by Eq. (8) within 65–290 K (solid curve) and extracted fitting parameters. Arrows point to the positions of the change of slope in the $v_Q(T)$ dependence at 210 K and at 65 K. Inset in **(a)** shows that the $v_Q(T)$ dependence can be described by two linear functions in the ranges 65–210 K and 210–290 K. For details, see Section V.2. **(b)** The temperature dependences of $^{63}$Cu(1) NQR line-widths for CuS phase in samples No.1 and No.5 (closed and opened circles, respectively). Arrow points to the temperature 55 K, below which strong broadening starts up. The solid line is the fit by the linear function within 55–290 K. For details, see Section V.2.

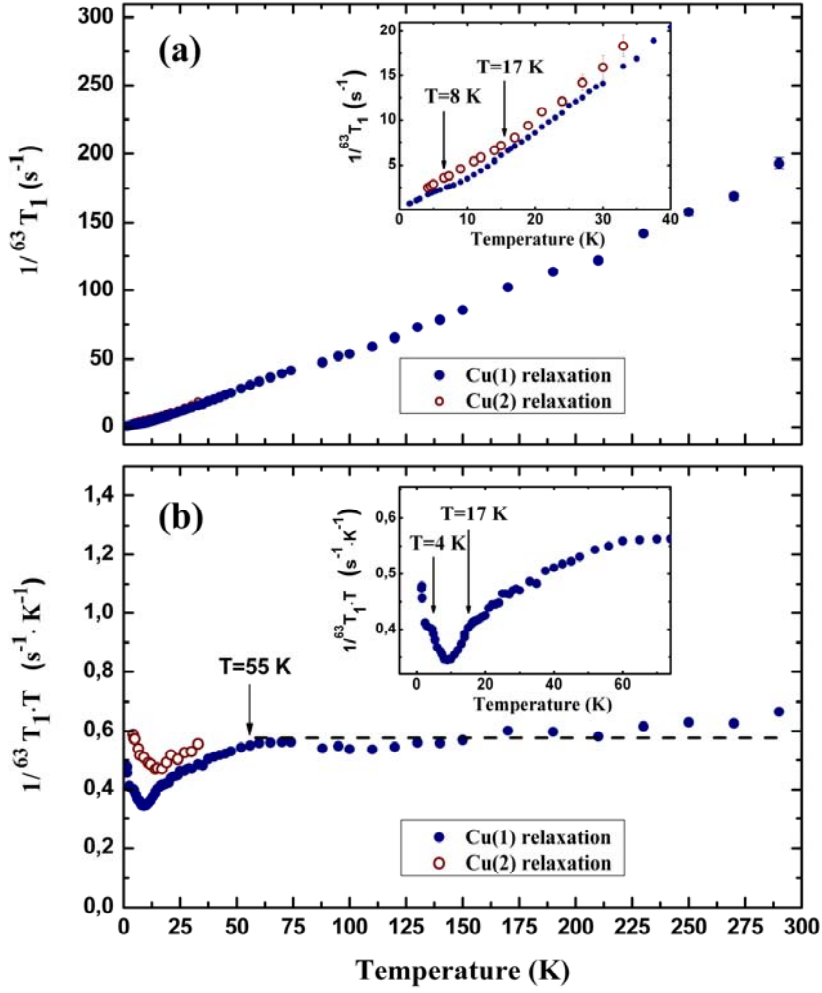

**FIG.5. (a)** The temperature dependence of nuclear spin-lattice relaxation $T_1^{-1}$ for $^{63}$Cu(1) for CuS phase in sample No.1 (closed circles). Inset in **(a)** shows the $T_1^{-1}(T)$ dependences for $^{63}$Cu(1) and $^{63}$Cu(2) within the range 1.47–40 K (closed and opened circles, respectively). Arrows point to the positions of bends in $T_1^{-1}(T)$ at about 17 K and 8 K. For details, see Section V.3. **(b)** The temperature dependences of $(T_1 \cdot T)^{-1}$ for $^{63}$Cu(1) and $^{63}$Cu(2) for CuS phase in sample No.1 (closed and opened circles, respectively). Arrow points to the temperature 55 K, above which the constant behavior of $(T_1 \cdot T)^{-1}$ holds approximately (dashed line) and below which the anomalous behavior of relaxation starts up. Inset in **(b)** shows the $(T_1 \cdot T)^{-1}$ dependence for $^{63}$Cu(1) within the range 1.47–74 K, arrows point to the positions of bends in $(T_1 \cdot T)^{-1}$ at about 17 K and 4 K. For details, see Sections V.3 and VI.3.

TABLE I. Interatomic distances and bond angles in covellite CuS above and below the temperature of phase transition ($T_{PT}$ = 55 K).

| | Hexagonal symmetry, 295 K | | | | Orthorhombic symmetry, 8 K |
|---|---|---|---|---|---|
| | Ref. 29 | Ref. 32 | Ref. 21 | Ref. 33 | Ref. 24 |
| Cu(1)-S(1)<br>-S(1) | 3×2,19 Å | 3×2,195(5) Å | 3×2,1905(2) Å | 3×2,1915(4) Å | 2×2,18(2) Å<br>1×2,17(2) |
| Cu(2)-S(1)<br>-S(2)<br>-S(2) | 1×2,34<br>3×2,30 | 1×2,334(6)<br>3×2,312(4) | 1×2,331(2)<br>3×2,305(2) | 1×2,339(2)<br>3×2,305(1) | 1×2,328(4)<br>1×2,32(2)<br>2×2,281(8) |
| S(2)-S(2) | 1×2,09 | 1×2,037(11) | 1×2,071(4) | 1×2,086(7) | 1×2,03(2) |
| Cu(1)-Cu(2) | 3×3,21 | | 3×3,199(4) | | 1×3,260(5)<br>2×3,044(7) |
| S(1)-S(2) | 3×3,75 | 3×3,773(9) | | 3×3,754* | 1×3,642*<br>2×3,793* |
| S(1)-Cu(2)-S(2) | 107° | 108,6(3)° | 108,16(7)° | 108,04(8)° | |
| S(2)-Cu(2)-S(2) | 111 | 110,3(3) | 110,76(7) | 110,87(8) | |
| S(1)-Cu(1)-S(1) | | 119,7(3) | | | |
| S(1)-S(1)-S(1) | | 120,0(0) | | | |
| Cu(2)-S(1)-Cu(2) | | | | | 170° |

* The values were calculated by Ref. 25 on the basis of data from corresponding references.

TABLE II. The magnitudes of asymmetry parameter $\eta$ in covellite CuS above and below the temperature of phase transition ($T_{PT}$ = 55 K).

| | $T > T_{PT}$ | | $T < T_{PT}$ | |
|---|---|---|---|---|
| | Ref. 27 | Ref. 15 | Ref. 27 | Ref. 15 |
| Triangular units [Cu(1)-S(1)$_3$], $\eta$, (a. u.) | ~ 0 (100 K) | – | – | – |
| Tetrahedral units [Cu(2)-S(2)$_3$S(1)], $\eta$, (a. u.) | ~ 0 (100 K) | 0? (60 K) | ≠ 0 and/or $v_Q$<<1.5 MHz (35 K) | 0.50–0.55 (15 K) |